\documentclass[]{spie}  
\usepackage[dvips]{graphicx}

\title{Monte Carlo Methods for X-ray Dispersive Spectrometers} 
\author{John R. Peterson\supit{a}, J. Garrett Jernigan\supit{b}, Steven M. Kahn\supit{a}
\skiplinehalf
\supit{a}Department of Physics, Columbia University, New York, NY 10027 \\
\supit{b}Space Sciences Laboratory, University of California, Berkeley, CA 94720 \\
}

\authorinfo{Further author information: send correspondence to jrpeters@astro.columbia.edu}
 
\begin{document} 
  \maketitle

\begin{abstract}
We discuss multivariate Monte Carlo methods appropriate for X-ray
dispersive spectrometers.  Dispersive spectrometers have many advantages
for high resolution spectroscopy in the X-ray band.  Analysis of data from
these instruments is complicated by the fact that the instrument response functions
are multi-dimensional and relatively few X-ray photons are detected from
astrophysical sources.  Monte Carlo methods are the natural solution to these
challenges, but techniques for their use are not well developed.
We describe a number of methods to produce a highly efficient and flexible
multivariate Monte Carlo.  These techniques include multi-dimensional response
interpolation and multi-dimensional event comparison.
We discuss how these methods have been extensively used in the XMM-Newton
Reflection Grating Spectrometer in-flight calibration program.  We also
show several examples of a Monte Carlo applied to observations of
clusters of galaxies and elliptical galaxies with the XMM-Newton observatory.
\end{abstract}

\keywords{Monte Carlo Techniques;  X-ray Spectroscopy;  Diffuse X-ray Sources;  X-ray Data Analysis}

\section*{Introduction}

The X-ray band ($\sim$ 0.1 to 10 keV or $\sim$ 1 to 100 $\mbox{\AA}$) contains the K and L shell transitions of abundant metals.  This allows the use of a wealth of spectral diagnostics in order to understand the physical conditions of highly ionized plasmas in astrophysical systems.  Historically, X-ray instruments have used both dispersive and non-dispersive spectrometers in order to resolve individual spectral features.  Non-dispersive spectrometers, such as proportional counters, CCDs, and calorimeters, directly measure the energy of the incident photon by dissipative processes.  In contrast, dispersive spectrometers such as gratings or crystals, measure the wavelength of the incident photons by recording the distance the incident photon was dispersed from a nominal focus.  Dispersive spectrometers have higher resolution at lower energies and non-dispersive spectrometers have higher resolution at higher energies.  Development of instruments utilizing either approach has proceeded in parallel and probably will continue for the forseeable future.

Both the {\it XMM-Newton} (Jansen et al.~\cite{jansen}) and {\it Chandra} (Weisskopf et al.~\cite{weisskopf}) observatories contain dispersive spectrometers, which now routinely resolve emission or absorption lines from individual ions.  Both observatories have collected hundreds of high resolution spectra of many different kinds of X-ray sources and have dramatically changed our physical understanding of many of these sources.
The Reflection Grating Spectrometers (den Herder et al.~\cite{denherder}) (RGS) on {\it XMM-Newton} consist of two arrays of reflection gratings placed behind nested mirror shells.  X-rays are dispersed to an array of CCDs.  {\it Chandra} has two arrays of transmission gratings, the low energy transmission grating (LETG) and the high energy transmission gratings (HETG).  The X-rays are also dispersed to either an array of CCDs or a micro-channel plate.  

A variety of new techniques to analyze data from these dispersive spectrometers are being developed which differ from many approaches previously used in X-ray astronomy.  In this presentation, we focus on a number of multivariate Monte Carlo techniques, which were developed for use with the Reflection Grating Spectrometers.  These methods have proved particularly useful both in the analysis of extended sources, such as galaxy clusters and elliptical galaxies, and as part of the in-flight calibration program.  These techniques have allowed the use of modeling approaches previously done with Monte Carlos, but rarely applied directly to the data.

\clearpage

\section*{The Nature of Data Sets Obtained with a Slitless Dispersive X-ray Spectrometer}

\begin{figure}
\begin{tabular}{c}
\includegraphics[height=10cm,angle=270]{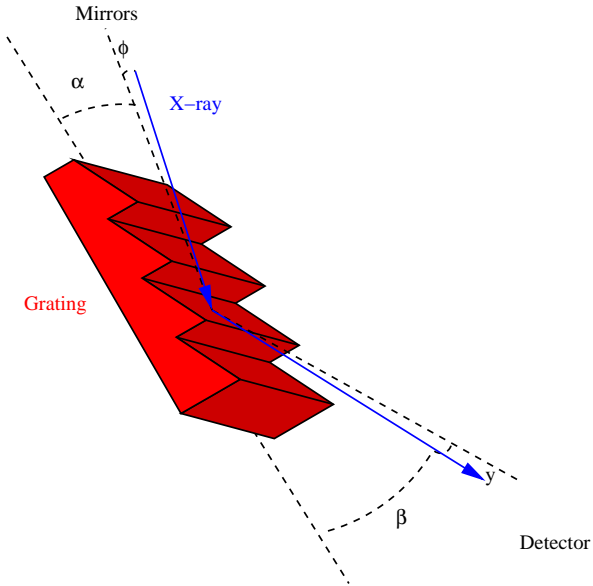}
\end{tabular}
\caption[grating]
{\label{fig:grating} Schematic of a reflection grating and the angles defined in the text.}
\end{figure}

Individual photons are collected with X-ray dispersive spectrometers.  Each photon's position on the detector ($\beta$, $y$), detector energy ($e$), and arrival time ($t$) are recorded.  We restrict attention to non-variable sources here, so we will ignore $t$.   The $\beta$-axis is aligned along the dispersion axis, and its value is then related to the wavelength of the incident photon by the dispersion equation.  The quantity $y$ is the off-axis direction perpendicular to the dispersion axis, which we will also refer to as cross-dispersion.  Photons dispersed by the grating obey the dispersion equation,

\begin{equation}
 \frac{m \lambda}{d} = \cos{\beta} - \cos{\alpha}.
\end{equation}

\begin{figure}
\includegraphics[height=17cm,angle=270]{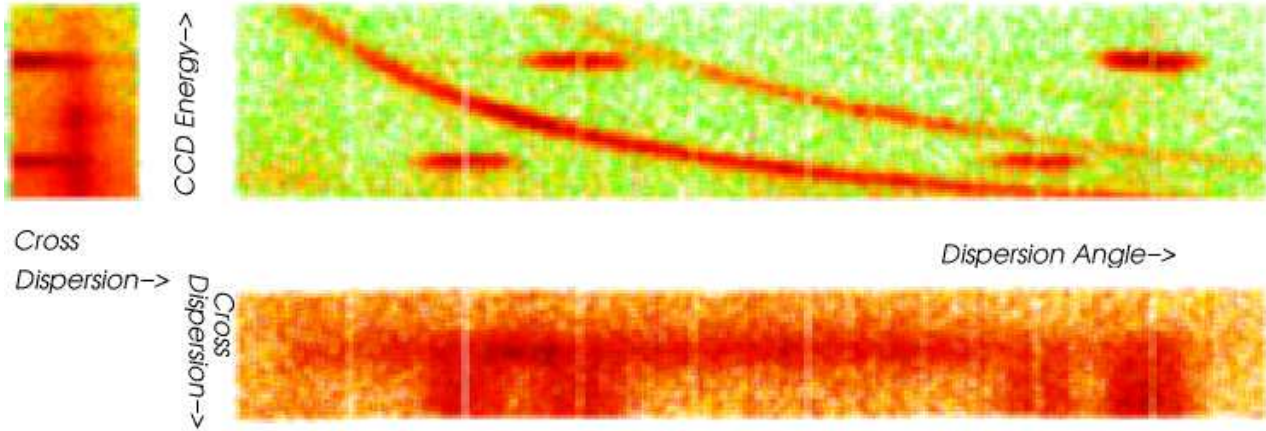}
\caption[s1101banana]
{\label{fig:s1101banana}  Raw RGS data for the galaxy cluster, S\'{e}rsic 159-03.  The plot consists of three panels for each of the two-dimensional projections of the three dimensional data.  The dispersion coordinate ($\beta$) vs. cross-dispersion coordinate ($y$) shows the dispersed spectral image.  It is blurred in the cross-dispersion direction due to the size of the source.  The three curved lines in the dispersion coordinate vs. CCD energy plot show the first, second, and third order dispersed spectra.  The four horizontal lines are the Al K and F K calibration sources.  Most of the photons in this image are due to Bremsstrahlung.   A darker region in the first order curved line is due to Fe L lines.
}
\end{figure}

\noindent
Here $d$ is the grating spacing, $\beta$ is the dispersion angle, $\alpha$ is the incident angle related to the direction of the source on the sky, and $m$ is an integer representing a spectral order. If the photon energy is crudely measured non-dispersively by the detector, and $\alpha$ is known, the measurement of $\beta$ give a unique determination of the photon wavelength at high resolution.  A diagram of the angles and a reflection grating is shown in Figure~\ref{fig:grating}.

\begin{figure}
\begin{tabular}{c}
\includegraphics[height=10cm]{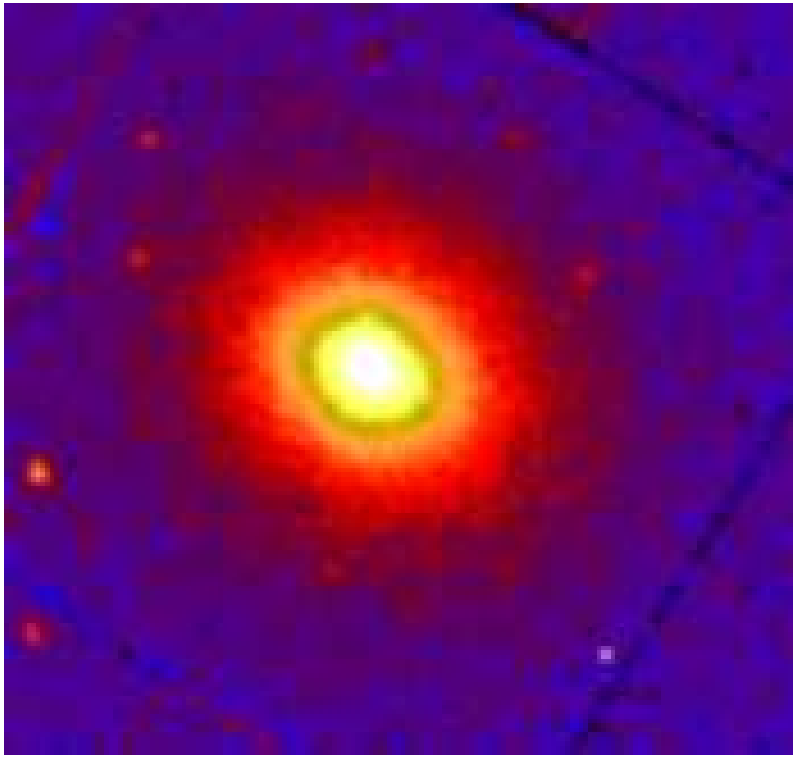}
\end{tabular}
\caption[s1101image]
{\label{fig:s1101image} X-ray Image of the galaxy cluster, S\'{e}rsic 159-03, using the EPIC-MOS detector.  This is the same source as in Figure~\ref{fig:s1101banana}.
}
\end{figure}

 For an extended source the situation gets more complex, as we discuss in detail below.  In that case, $\alpha$ is not uniquely known, causing an ambiguity in the ($\beta$, $\lambda$) relation.  The quantity $y$ then also contains useful information about the spatial distribution perpendicular to the dispersion axis.  An example of the data of an RGS observation of a galaxy cluster, S\'{e}rsic 159-03, is shown in Figure~\ref{fig:s1101banana}.  The corresponding image for a non-dispersive spectrometer is seen in Figure~\ref{fig:s1101image}.  It is immediately apparent that what is seen in a dispersive spectrometer is very different than what is seen on the sky.

Typically, observations contain between $10^3$ and $10^6$ photons.  All three measurements ($\beta$, $y$, $e$) contain useful information in the case of an extended source.  The 3-dimensional data space is also always sparsely filled.  This means the number of photons is always less than the number of effective resolution elements ($10^7$ to $10^9$).  Clearly, this sparse multi-dimensional data is common in other applications.

\section*{Instrument Response Calculation}

The standard method of response calculation in X-ray astronomy is to convolve the model for the spectrum, S, with a response kernel, R (see e.g. Arnaud~\cite{Arnaud}).  This can be expressed as

\begin{equation}
 P (\beta) = \int ~d\lambda ~R (\beta | \lambda) ~S ( \lambda )
\end{equation}

\begin{figure}
\includegraphics[height=3cm]{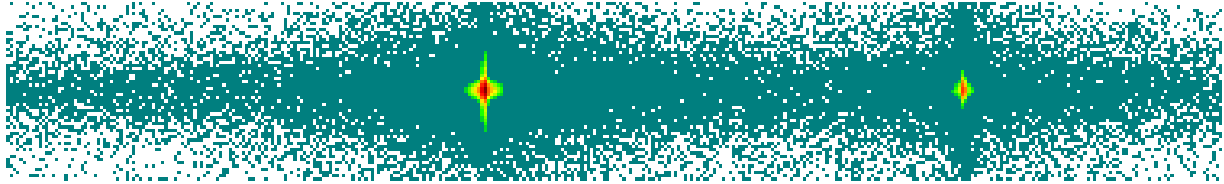}
\caption[ptimage]
{\label{fig:ptimage}
Simulation of the RGS with a monochromatic point source.  The vertical axis is cross-dispersion and the horizontal axis is the dispersion angle.  The two spots correspond to the first and second order point.  The large angle scattering wings can be seen in both direction.  The distribution depends on the incidence angle, energy, and dispersion angle.
}
\end{figure}

\begin{figure}
\includegraphics[height=2.7cm]{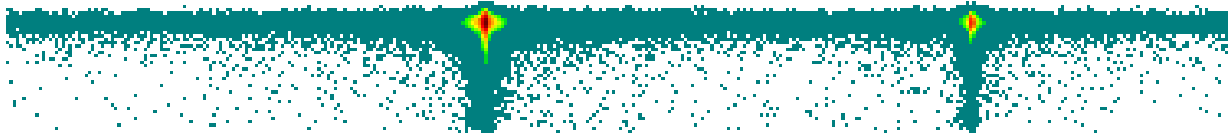}
\caption[ptbanana]
{\label{fig:ptbanana}
Simulation of the RGS with a monochromatic point source.  This is the same as Figure~\ref{fig:ptimage} but plotting the CCD energy vs. the dispersion angle.  The scattering wings are still visible at the horizontal line and low pulseheight events are part of a partial event tail.
}
\end{figure}

\noindent
Here P is the probablility of a photon being detected with a given wavelength, R is the response kernel, and S is the input spectral model.  It should be noted that R and S are rarely analytic functions.   The integral is normally computed numerically by converting the above expression into a sum.  P is then compared directly with the detected data histogram.  Figures~\ref{fig:ptimage} and~\ref{fig:ptbanana} show the RGS response functions for an on-axis monochromatic source. 

When we extend this calculation to the case of an extended source, it becomes a three dimensional integral. 

\begin{equation}
 P (\beta, y, e) = \int ~d \alpha ~d \phi ~d \lambda ~R (\beta , y, e | \lambda, \alpha, \phi) ~S ( \lambda, \alpha, \phi )
\end{equation}

Calculating the three dimensional integral is impractical, since we must integrate over the sky coordinates ($\alpha$ and $\phi$) in addition to the wavelength.  This is especially true if one wants to modify the model function, S, iteratively.  Often model fitting involves trying many different models and many different parameters.  Some approximations can be made, such as assuming the source's spectrum does not vary as a function of spatial position and assuming the response function does not change too much off-axis.  These are useful in some circumstances, but the general case requires a Monte Carlo.  Then each photon can be chosen from a distribution of S with a $\lambda$, $\alpha$, and $\phi$.  This then can be converted to a $\beta$, $y$ and $e$ by using probability response functions.  In this way, the probabilities of all possible $\beta$, $y$, and $e$ do not have to be calculated at every $\lambda$, $\alpha$, and $\phi$.
Below we focus on two sets of challenges that had to be overcome in order to use Monte Carlos for efficient analysis of the data. 

\section*{Optimization of the Response Calculation}

The calculation of a single element of the RGS response is computationally intensive. This is due to the fact that many components of the response, such as the effect of the mirror, grating, and detector, have to be convolved in the calculation.  Each usually has its own energy and spatial dependence.  For this reason it is extremely useful to save the response probabilities in pre-calculated tables.  The probabilities can be looked up for each photon depending on its sky position and wavelength ($\alpha$, $\phi$, and $\lambda$).  This then decreases the computation time by factors of $10^4$ or more.

\subsection*{Monte Carlo Interpolation:  Morphing the Probability Functions}

\begin{figure}
\includegraphics[height=15cm,angle=270]{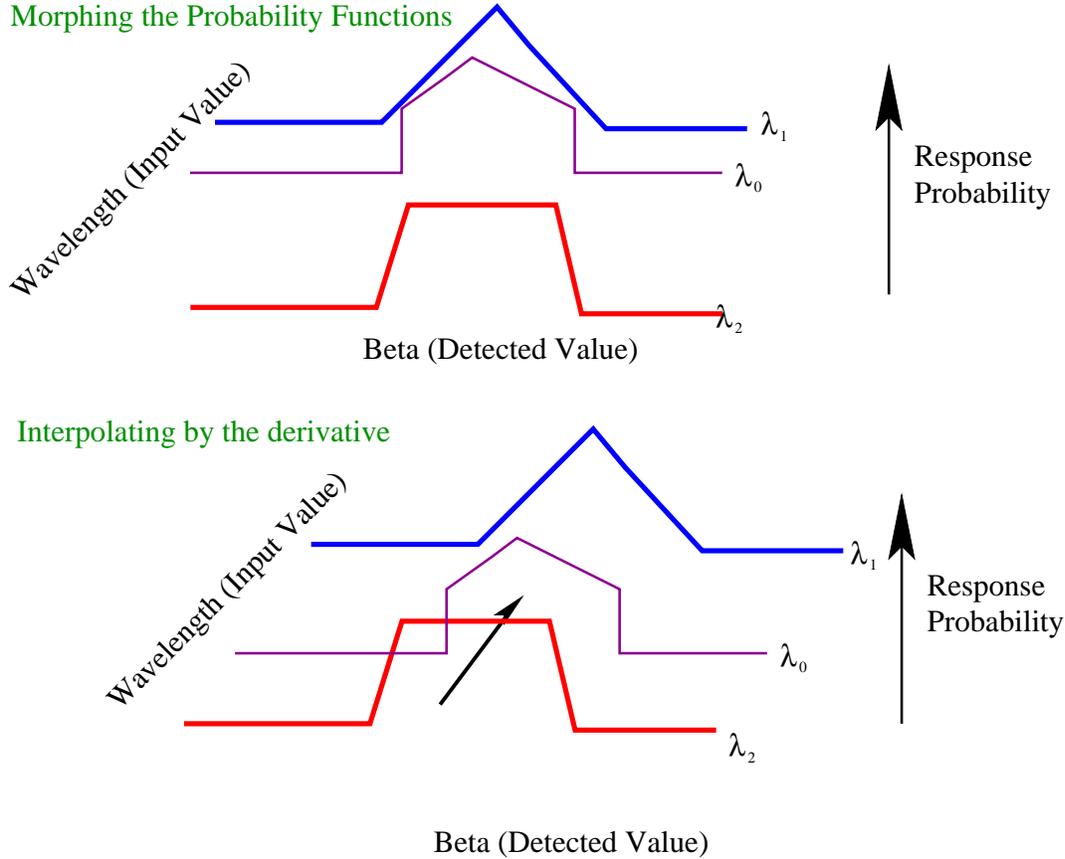}
\caption[morph]
{\label{fig:morph} The two steps of looking up the instrument response, which is saved on coarse grids.  The first panel shows Monte Carlo morphing and the second panel shows Monte Carlo interpolating.  An explanation of the method is in the text.}
\end{figure}

An immediate consequence of this, however, is that the multi-dimensional tables if saved at high enough resolution take up too much memory.  For two dimensional tables this would not be the case.  For the response for the RGS, however, the six dimensional function, $R$, could only be broken into a series of two or three dimensional functions.  These functions can only be saved in tables with coarse grids.  

A solution to using the coarse grids, however, is to approximate the probability density functions between grid points as a linear combination of the response at the grid points.  With a Monte Carlo, this is accomplished by the following method.  Consider that the probability of a photon having a given $\beta$ for a given $\lambda$, $P(\beta| \lambda)$, is calculated at two wavelengths, $\lambda_1$ and $\lambda_2$ as shown in the top panel in Figure~\ref{fig:morph}.  If we want to know the response at some $\lambda_0$ (such that $\lambda_1<\lambda_0<\lambda_2$), the response function at $\lambda_1$ can be used some fraction of the time and the response function at $\lambda_2$ can be used the other fraction of the time.  The fraction is determined by the distance $\lambda_0$ is from $\lambda_1$ and $\lambda_2$.

\subsection*{Monte Carlo Interpolation:  Shifting by the Derivative}

An additional step is needed if the response functions have sharp peaks in them that shift as a function of $\beta$.  We then shift the output value, $\beta$, by the first order behaviour of the response.   Say we use the probability distibution at $\lambda_2$ and choose a $\beta$=$\beta_2$ based on the distribution, P($\beta | \lambda_2$).  The final value of $\beta$ is given by $\beta_2+\frac{d \beta}{d \lambda}|_{\lambda_2}(\lambda_0-\lambda_2)$.  $\frac{d \beta}{d \lambda}$ is calculated by the dispersion equation.  This method causes the distribution of $\beta$ to approach the probability distribution shown in the second panel in Figure~\ref{fig:morph}.  The combined implementation of these two techniques results in photons being created at 50,000 events per second per GHz of processor speed and can be generalized to more dimensions.

\section*{General Method}

\begin{figure}
\includegraphics[height=18cm]{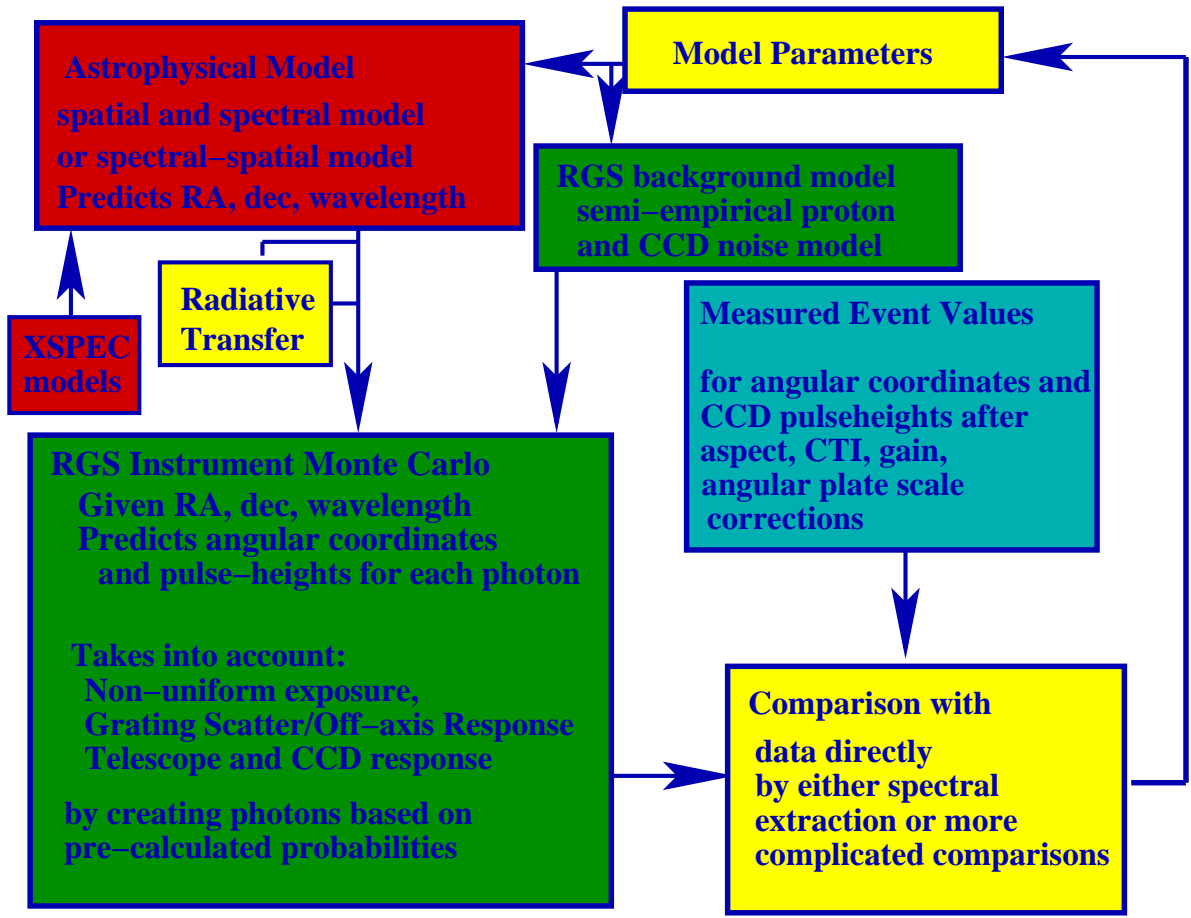}
\caption[flowchart]
{\label{fig:flowchart} 
Basic flowchart for the method of model comparison presented in this paper.
}
\end{figure}

Once this Monte Carlo is built, individual models can be used and then compared to the data.  A procedure similar to that described in the flowchart in Figure~\ref{fig:flowchart} can be adopted.

\begin{itemize}

\item{First, models are formulated in terms of a set of parameters.  These parameters predict probability distributions for the spectral and spatial models.}

\item{Photons are then drawn from these probability distributions and assigned a wavelength and two sky coordinates. ($\lambda$, $\alpha$, $\phi$).}

\item{The dispersion, cross-dispersion, and pulseheight values ($\beta$, $y$, $e$) are predicted from the instrument Monte Carlo.  In addition, a background Monte Carlo predicts background events with $\beta$, $y$, and $e$ values.}

\item{Finally, the simulated events are compared with the measured photon events in a variety of ways.  Some of these methods are described in the next section.  Parameters are modified to improve the consistency of the data with the model.}

\end{itemize}

\begin{figure}
\includegraphics[height=18cm]{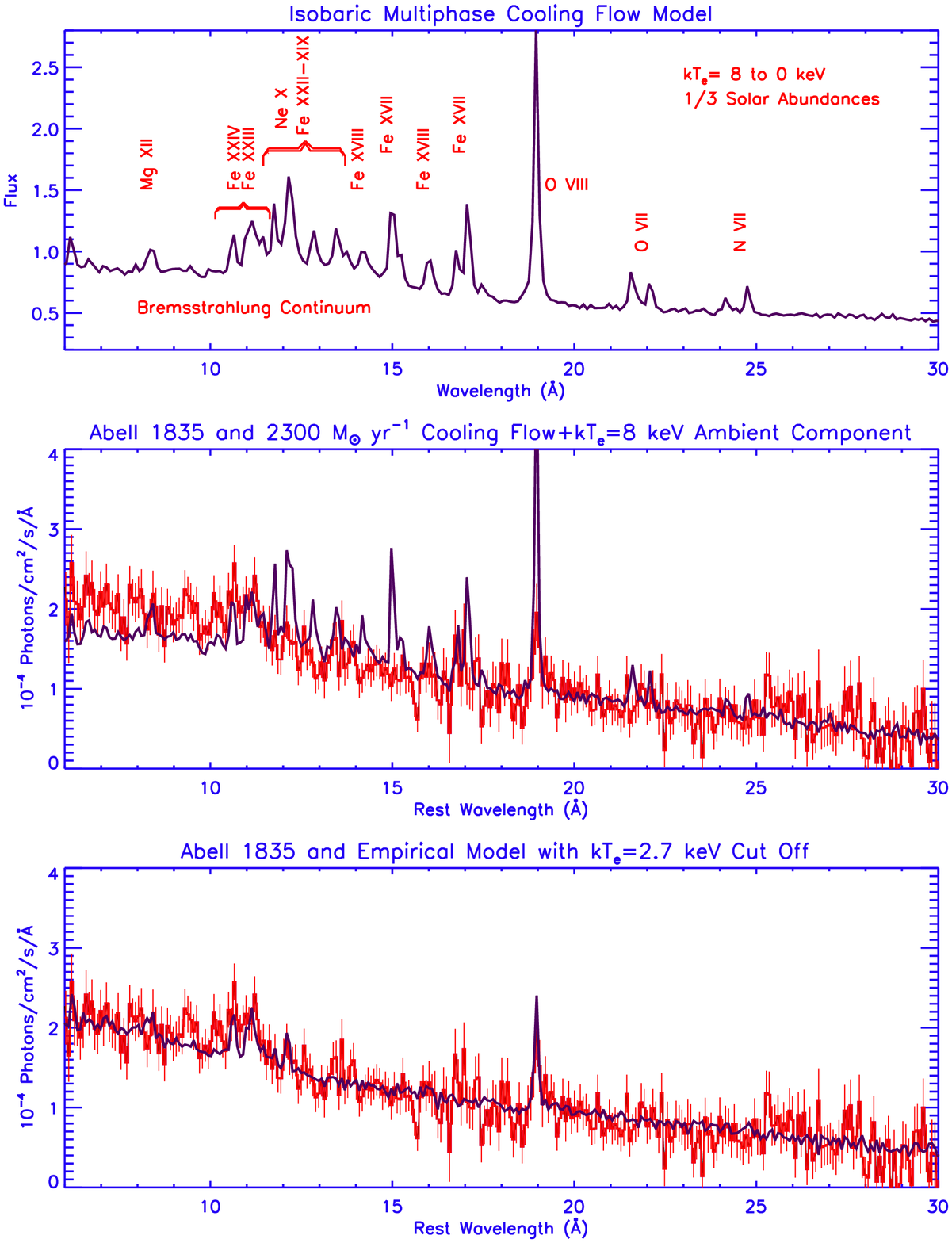}
\caption[cflow]
{\label{fig:cflow} The prediction of the isobaric cooling flow model and comparison with the massive cooling-flow galaxy cluster, Abell 1835.  The top panel shows the spectral prediction for cooling-flows.  The middle panel shows the model (blue) compared with the observed data (red).  The bottom panel shows an empirical model with the emission below 2.7 keV suppressed.}
\end{figure}

\section*{Event Comparison}

There are a variety of methods to compare one set of detected events to another set of simulated events.  

\subsection*{Transforming and Cutting the Data}

  A major advantage of the Monte Carlo approach, which has been recognized for some time, is that the simulated events can be selected and transformed in various ways in methods identical to what is done with the real data.  There is then no need to worry about biasing the data or misinterpreting the detection, because a Monte Carlo already explicitly takes into account these data analysis steps.

In our case, the transformation of the data into a binned spectrum takes several steps.  For an extended source, the assignment of wavelengths is not unique.  The effect of various data selection cuts is difficult to correct for by any other method than an explicit Monte Carlo.  
With the Monte Carlo, however, wavelength can be assigned by fixing the incidence angle, $\alpha$, in the same way for the simulated photons and the detected photons.  The normalization is properly calculated by the direct sampling Monte Carlo method.  A two-sample $\chi^2$ statistic can be constructed to compare the model with the data.  If $V_{1j}$ is the observed data values in the $j$th bin, and $V_{2j}$ is the observed model values and there are $n$ events in the data and $m$ in the model, then the $\chi^2$ statistic is given by

\begin{equation}
\chi^2=\sum_j \frac{|V_{1j}-\frac{n}{m} V_{2j}|^2}{\frac{n}{m}(V_{1j}+V_{2j})} .
\end{equation}

 This method has been applied in several cases.  An important result is shown in Figure~\ref{fig:cflow} in the study of cooling-flows in massive clusters of galaxies (Peterson et al.~\cite{Peterson}).  Here it has been shown that the simple thermodynamic prediction for the soft X-ray spectrum of cooling-flows fails to reproduce the observed data.

The model consists of the following.  An isothermal envelope of hot plasma is surrounds a three dimensional cooling region with a different X-ray spectrum.  The second X-ray spectrum consists of the unique model for an optically-thin multiphase isobaric cooling plasma.  Photons are chosen based on the plasma emissivity and then projected on the sky.
An important aspect of the Monte Carlo method in this case was that the straight-forward theoretical description of the cluster had different spectra at each spatial position.

\subsection*{Multi-dimensional Cram\'{e}r-von Mises}

Instead of explicitly extracting a subset of photons and assigning a wavelength based on the dispersion equation, a multi-dimensional event comparison statistic can be used.  This is particularly important in estimating the background contribution to the detected events.  The Cram\'{e}r-von Mises statistic (Anderson and Darling~\cite{Anderson}) is calculated in a similar way to the more familiar Kolmogorov-Smirnov statistic.
It is determined by computing the cumulative distribution, $F_1$ of the $n$ data values and the cumulative distribution, $F_2$ of the $m$ model values and then evaluating the
following sum at each of the model and data values.

\begin{equation}
 W^2 = \frac{m n}{{\left( m + n \right)}^2}
\sum_{x={\rm{data~and~model~values~of~x}}} {\left[ F_1(x) - F_2(x) \right]}^2 
\end{equation}

\begin{figure}
\includegraphics[height=18cm]{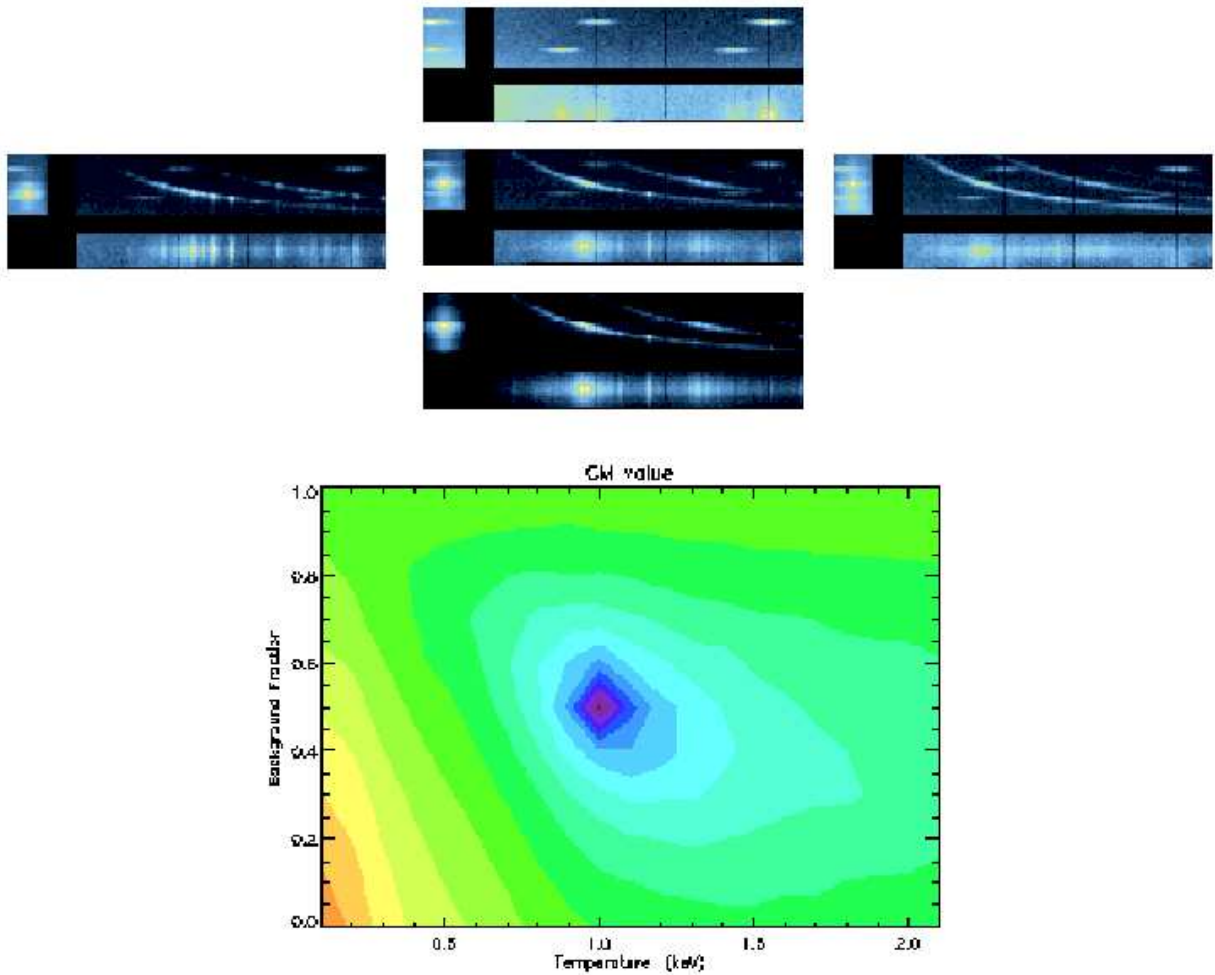}
\caption[param]
{\label{fig:param}  An example of model iteration using the Cram\'{e}r-von Mises multivariate statistic.  The top five plots show the projections of the data for five separate simulation.  Each set of three plots has the same meaning as the plots in Figure~\ref{fig:s1101banana}.  The center simulation is a 1 keV isothermal beta model of a cluster of galaxies plus an instrument background model.  Top and bottom plots are the same simulation but with a larger and smaller, respectively, background fraction.  The left and right plots are 0.5 and 2.0 keV simulations.  The bottom countour plot shows the value of the Cram\'{e}r-von Mises statistic for various combinations of the temperature and the background fraction of the model like those shown in the simulation above.  It is compared to the simulation in the center plot.  The contour plot then demonstrates the the statistic is capable of finding the right solution.
}
\end{figure}

This works for one-dimensional data and is useful for sparse data.  The multi-dimensional analog of this statistic is obtained by using a linear combination of the measured values ($\beta$, $y$, and $e$) in each dimension, and using that for the value of $x$ in the above equation.  Various linear combinations can be used to emphasize one dimension over another, and it appears that summing over the various combinations results in a robust multi-dimensional statistic.  A series of simulations in Figure~\ref{fig:param} demonstrates the use of this statistic.

\subsection*{Emission Line Profiles}

\begin{figure}
\includegraphics[height=15cm]{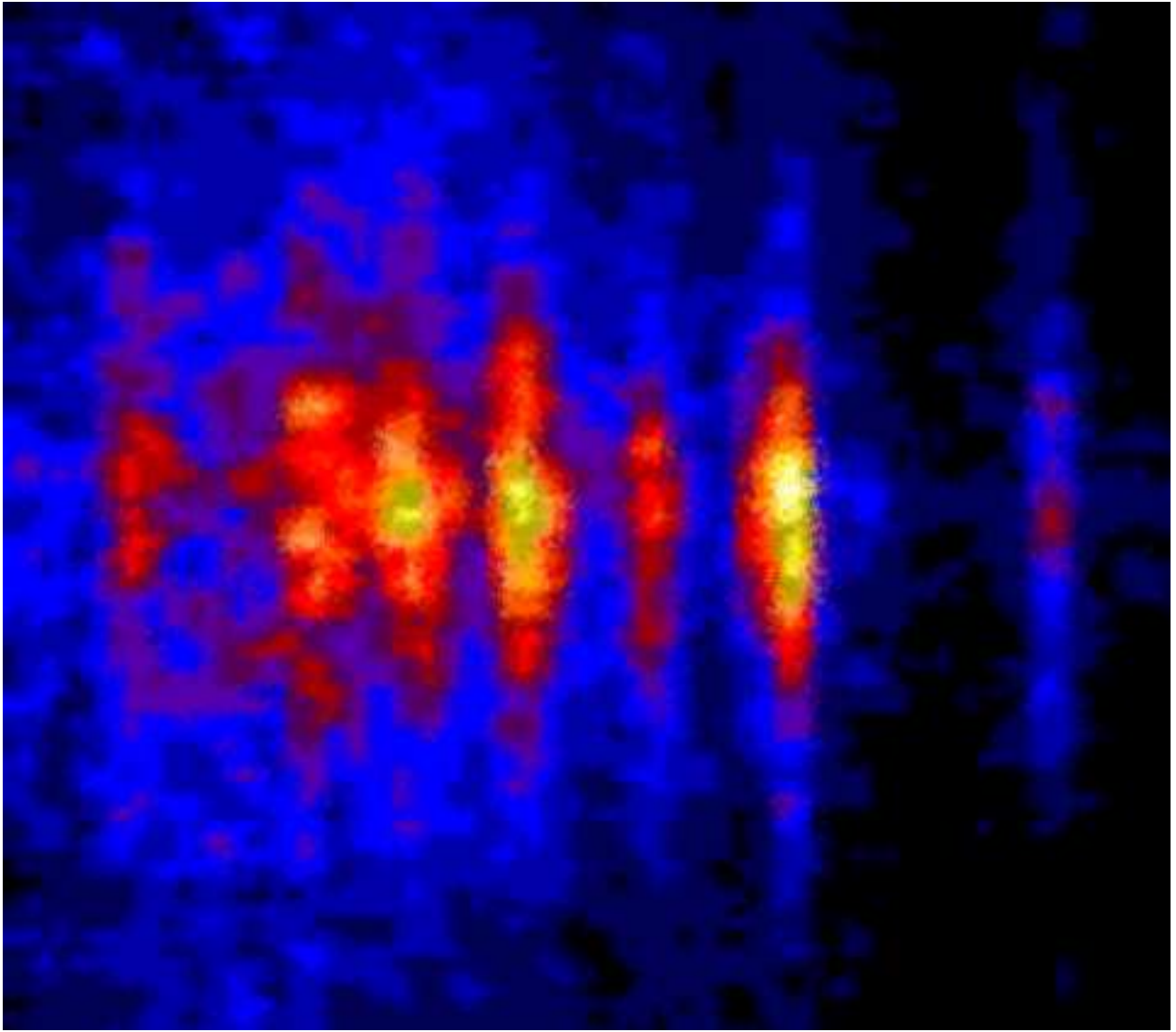}
\caption[ngc4636image]
{\label{fig:ngc4636image}  Cross-dispersion vs. dispersion angle images of emission lines of the elliptical galaxy, NGC 4636.  The vertical axis is the spatial distribution in the cross-dispersion axis in each emission line; whereas, the horizontal axis is the wavelength.  The images are (from left to right), Ne X, Ne IX, Fe XVIII 2p-3d, Fe XVII 2p-3d, Fe XVIII 2p-3s, Fe XVII 2p-3s, and O VIII.
}
\end{figure}

\begin{figure}
\includegraphics[height=16cm,angle=270]{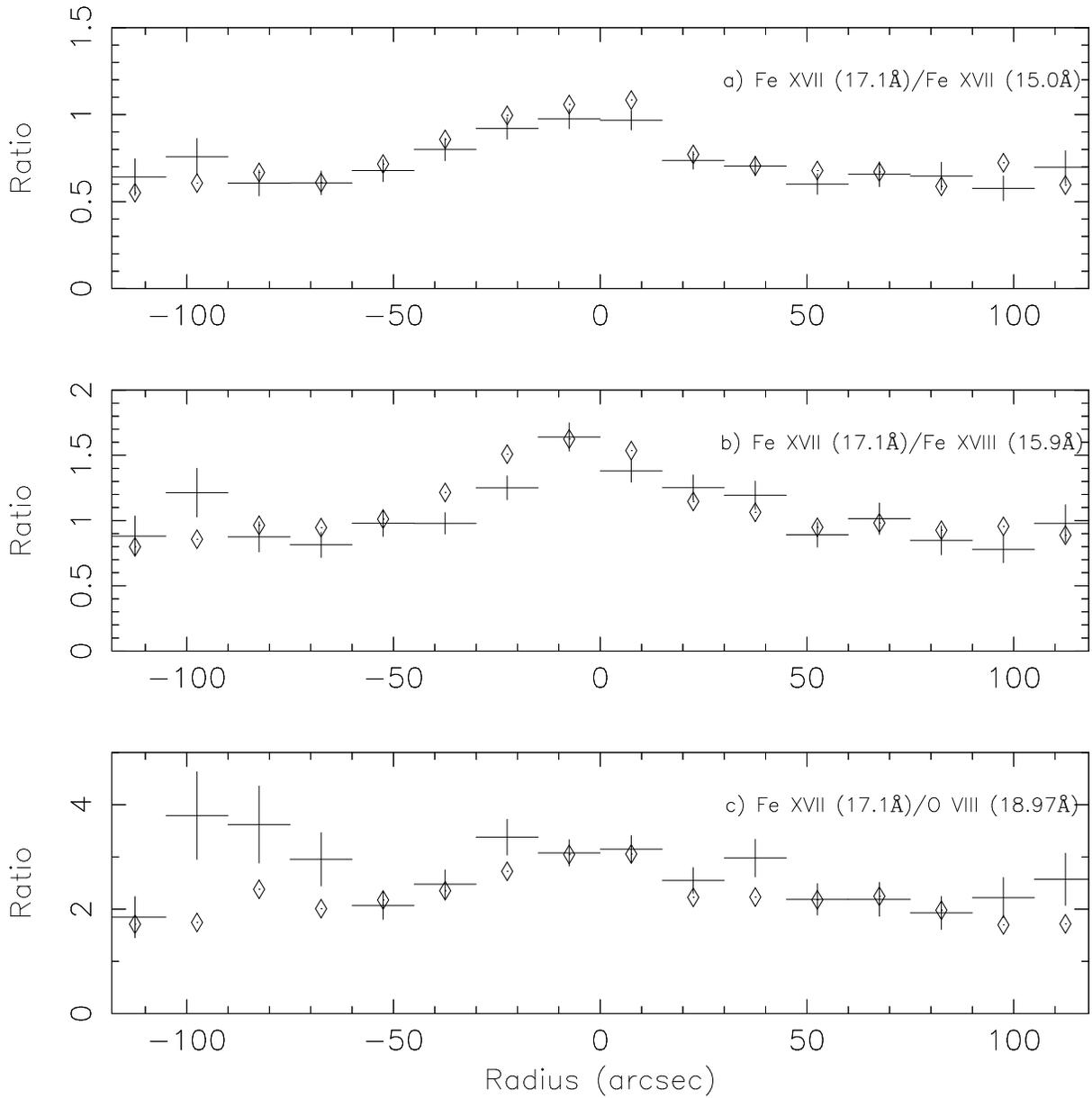}
\caption[ngc4636xdsp]
{ \label{fig:ngc4636xdsp} Ratios of the number of photons as a function of cross-dispersion angle for various emission lines.  This plot is made by selecting narrow wavelength regions (horizontal axis) in Figure~\ref{fig:ngc4636image} and then plotting the ratio of the number of counts.  The same thing is done for a simulation with a temperature and density distribution and resonant scattering Monte Carlo.
}
\end{figure}

A final method of comparison involves explicitly looking at the cross-dispersion profile of an individual emission line.  Again, the precise interpretation of such a distribution is difficult without using a Monte Carlo to select the events and simulate the effect of the data selection.  This is seen in the analysis of the elliptical galaxy, NGC 4636 (Xu et al.~\cite{Xu}).   Individual compressed emission line images can be seen in Figure~\ref{fig:ngc4636image}. Differences in the spatial distribution of certain emission lines can be used to provide joint constraints on physical properities of the source.  Figure~\ref{fig:ngc4636xdsp} shows three ratios of emission lines in the cross-dispersion direction.  The first panel is an indication that the X-ray plasma is cooler in the center of the elliptical galaxy.  The second panel is an indication that one emission line is optically thick and photons have been redistributed on the sky due to resonant line scattering.  The third panel indicates that oxygen and iron have a similar spatial distribution.  This yields joint contraints on the density and temperature structure and provides robust determinations of the abundances of various metals.

\section*{Future Work}

An undesirable aspect of this method is that convergence of the model is complicated by the fact that the model has statistical noise in it as well.  This is partly overcome by simulating far more events in the model than are detected in the data.  Additionally, careful consideration of what parameters affect what parts of the data allows the vast parameter space to be more efficiently searched.  We expect new methods, however, could develop this further.

\end{document}